\documentstyle[12pt]{article}

\begin{document}
\begin{titlepage}
\begin{center}
{\LARGE \bf The Lagrangian of self-dual  \\[.3em]

gravitational field as a limit of \\[.3em]

the SDYM Lagrangian \\[.3em]}

\vspace{2cm}
{\bf Jerzy F. Pleba\'nski\footnote{{}{E-mail address: pleban@fis.cinvestav.mx}}
and Maciej Przanowski\footnote{{}{Permanent 
address: Institute of Physics, Technical University of L\'od\'z, 
W\'olcza\'nska 219, 93-005 L\'od\'z, Poland }} }
\\
{\it Department of Physics \\
Centro de Investigaci\'on y de Estudios Avanzados del IPN \\ Apartado Postal 
14-740, M\'exico 07000, D.F., M\'exico\\}

\vspace{2.5cm}
{\Large \bf Abstract \\}
\end{center}
The action for the su(N) SDYM equations is shown to give in the limit     
$N \to \infty$ the action for the sixdimensional version of the second 
heavenly equation.  The symmetry reductions of this latter equation to the 
well known equations of self-dual gravity are given.  The Moyal deformation of 
the heavenly equations are also considered.\\

\end{titlepage}

Recently, a great deal of interest has been devoted to symmetry reductions 
of the SDYM equations to integrable equations of mathematical physics [1--16]. 
In particular many works concern the symmetry reductions of the SDYM equations 
to the self-dual gravity equations [6--16].

In our previous works we have found the general form of the su(N) SDYM 
equations which in the limit N $\to \infty$ gives the sdiff $(\Sigma^2)
\cong \ su \ (\infty)$ SDYM equation in the form the second heavenly 
equation in six dimensions [17, 18].  Then we have shown that the symmetry 
reductions of that equation lead to the well known heavenly equations of 
self-dual gravity.

We were also able to obtain the natural Moyal deformation of the 
heavenly equations. This deformation has been previously found by Strachan 
[12, 19] and Takasaki [20].

In the present work we are going to show that the limiting process 
$N \to \infty$ leading from su(N) SDYM equations to the heavenly equations 
can be also well defined on the action level.  Similar considerations enable 
one to find the action for the Moyal deformation of the {\it sixdimensional 
version of the second heavenly equation}.

We deal with the su(N) SDYM equations in the flat 4-dimensional real, 
simply connected flat manifold $V \subset R^4$ of the metric

\begin{equation}
ds^2 = 2 (d x \otimes_s d \tilde x + dy \otimes_s d\tilde y) 
\end{equation}
where $x, y,\tilde x, \tilde y$ are null coordinates on $V$ and $\otimes_s$ 
denotes the symmetrized tensor product, i.e.,  $d x \otimes_s d \tilde x 
=  \frac{1}{2} (d x \otimes d \tilde x + d \tilde x \otimes d x)$. The 
coordinates $(x,y, \tilde x, \tilde y)$ are chosen in such a manner that 
the su(N) SDYM equations read [21]

\begin{eqnarray}
F_{xy}  &= &0  \\ 
F_{\tilde x \tilde y}  &= &0   \\
F_{x \tilde x} \ + \ F_{y \tilde y} &= &0, 
\end{eqnarray}
where, as usually, $F_{\mu \nu} \ \in  \mbox{su (N)} \otimes C^\infty
(V), \mu , \nu  \in  \{x, y, \tilde x, \tilde y\}$, stands for the 
Yang-Mills field tensor. 

Then, in terms of the Yang-Mills potentials $A_\mu \in \mbox{su (N)} 
\otimes C^\infty (V)$ which define $F_{\mu \nu}$ according to the well 
known formula.

\begin{equation}
F_{\mu \nu} = [\partial_\mu + A_\mu , \partial_\nu + A_\nu ]
\end{equation}
one rewrites the equations (2.a, b, c) as follows 

\begin{eqnarray}
\partial_x A_y - \partial_y A_x + [A_{x,} A_y] &= &0\\ 
\partial_{\tilde x} A_{\tilde y} - \partial_{\tilde y} A_{\tilde x} + 
[A_{\tilde x,} A_{\tilde y} ] &= &0 \\ 
\partial_x A_{\tilde x} - \partial_{\tilde x} A_x + \partial_y A_{\tilde y} 
- \partial_{\tilde y} A_y + [A_{x,} A_{\tilde x} ] + [A_{y,} A_{\tilde y}] 
&= &0
\end{eqnarray}
 
Eq.~(4.a) implies that the potentials $A_x$ and $A_y$ are of the pure 
gauge form i.e.,\ there exists an SU(N)-valued function $g$ such that 

\begin{equation}
A_x = g^{-1} \partial_x g \ \ \ \mbox{and} \ \ A_y = g^{-1} \partial_y g.
\end{equation}

Therefore one can choose the gauge such that    

\begin{equation}
A_x = 0 \ \ \ \mbox{and} \ \ \ A_y = 0 
\end{equation}

\vskip.7cm
\noindent
Henceforth we assume that this last condition holds. Consequently, Eqs. (4.a, 
b, c), under the condition (6), read 

\begin{eqnarray}
\partial_{\tilde x} A_{\tilde y} - \partial_{\tilde y} A_{\tilde x} + 
[A_{\tilde x,} A_{\tilde y} ] &= &0\\ 
\partial_x A_{\tilde x} - \partial_y A_{\tilde y} &= &0
\end{eqnarray}

From (7.b) it follws that there exists an su(N)-valued function $\theta$ such 
that  

\begin{equation}
A_{\tilde x} = - \partial_y \theta \ \ \mbox{and} \ \ A_{\tilde y} = 
\partial_x \theta .
\end{equation}

Substituting (8) into (7.a) one gets   
\begin{equation}
\partial_x \partial_{\tilde x} \theta + \partial_y \partial_{\tilde y} 
\theta + [\partial_x \theta , \partial_y \theta ] = 0 ,
\end{equation}

\[
\theta \ \in  \ \ su(N) \ \otimes C^\infty (V). 
\]

Eq. (9) is equivalent to the su(N) SDYM equations (2.a, b, c).

Now, straightforward calculations show that Eq. (9) can be derived from the 
following least action principle

\[              
\delta S = 0, \ \ \ S = \int_V {\cal L} d v , \ \ \ d v = d x dy d \tilde x 
d \tilde y
\] 

\begin{equation}
{\cal L}: = \frac{(2\pi)^4}{N^3} Tr \ \{ \frac{1}{3} \theta [ \partial_x 
\theta , \partial_y 
\theta ]  - \frac{1}{2} \ \big( (\partial_x \theta) (\partial_{\tilde x}
\theta) + (\partial_y \theta)(\partial_{\tilde y} \theta) \big) \}
\end{equation}

Thus the Lagrangian ${\cal L}$ defined by (10) can be considered to be 
the Lagrangian for the SDYM field.

Now we let N tend to infinity.

Thus we arrive at the su$(\infty)$ algebra.  It is well known that [22-26]

\begin{equation} 
su \ (\infty) \cong \ \mbox{sdiff} \ (\Sigma^2) \cong \mbox{the Poisson algebra  on }\Sigma ^2 
\end{equation}
where $\Sigma ^2$ is a 2-dimensional real manifold. 

Employing the results of Refs [23, 24], where the case of $\Sigma^2$ being 
the 2-torus has been considered, one can quickly find that in order to 
obtain the $N \to \infty$ limit of the action (10) we can proceed as 
follows: \ We consider $\theta$ to be the function on $V \times 
\Sigma^2$ i.e., $\theta = \theta (x, y, \tilde x, \tilde y, p, q)$ where 
$(p, q)$ are the coordinates on $\Sigma^2$. Moreover, we make the following 
substitutions

\[
[ \cdot , \cdot ] \ \ \to \ \ \{ \cdot , \cdot \}_P
\]
\begin{equation}
\frac{(2\pi)^4}{N^3} T_r (\ldots ) \ \ \ \to \ \ \ - \int_{\Sigma^2} 
(\ldots ) dpdq, 
\end{equation}
where $\{ \cdot , \cdot \}_P$ stands for the Poisson bracket          

\begin{equation}
\{ f_1, f_2\}_P := \frac{\partial f_1}{\partial q} 
\frac{\partial f_2}{\partial p} \ - \ \frac{\partial f_1}{\partial p} 
\frac{\partial f_2}{\partial q} 
\end{equation}
for any $f_1 = f_1 (x, y, \tilde x, \tilde y, p, q)$ and $ f_2 = f_2 (x, y, 
\tilde x, \tilde y, p, q)$.

Thus the action $S$ defined by (10) is brought to the following form

\[
S \to S_\infty = \int_{V\times \Sigma^2} \  \bigg\{ - \frac{1}{3} \theta 
\{\partial_x \theta , \partial_y \theta \}_P + \frac{1}{2} \ ( (\partial_x 
\theta) (\partial_{\tilde x} \theta) 
\]

\begin{equation}
+ (\partial_y \theta)(\partial_{\tilde y} \theta)) \} d vdpdq.
\end{equation}

This shows that we now deal with the Lagrangian in a 6-dimensional space 
$V \times \Sigma^2$

\begin{equation}
{\cal L}_\infty = - \frac{1}{3} \theta \{ \partial_x \theta , \partial_y 
\theta\}_P + \frac{1}{2} ((\partial_x \theta)(\partial_{\tilde x} \theta) 
+ (\partial_y \theta)(\partial_{\tilde y} \theta))
\end{equation}

The similar Lagrangian in four dimensions was first considered in [31].

It is an easy matter to prove that the Euler-Lagrange equation for ${\cal L}_
\infty$ reads

\begin{equation}
\theta_{x \tilde x} + \theta_{y \tilde y} + \{ \theta_x, \theta_y\}_P = 0, 
\end{equation}
where we use the obvious notation $\theta_x: = \partial_x \theta , \theta_{
x \tilde x} := \partial_x \partial_{\tilde x} \theta , \ldots , etc.$

Eq. (16) resembles very much the well known second heavenly equation [27] and 
we call it the {\it sixdimensional version of the second heavenly equation}.

This equation has been found in our previous works [17, 18] as the result of 
the  $N \to \infty$ limit of Eqs. (7. a, b).

Here we show that the sixdimensional version of the second heavenly equation 
can be considered to be the $N \to \infty$ limit of the su(N) SDYM 
equations also on the action level.

Now we intend to present how the appropriate symmetry reductions of Eq. (16) 
lead to the heavenly equations (compare with [17, 18]).
 
(a) {\it The first heavenly equation}.
\noindent

Let

\[\theta = \theta ' - \frac{1}{2} (x \tilde x + y \tilde y)\]
\begin{equation}
\theta '_{x} = \theta '_{\tilde y}, \ \ \  \theta '_y = - \theta '_{\tilde x}.
\end{equation}

Then, the function $\theta '$ is of the form 

\begin{equation}
\theta ' (x, y, \tilde x, \tilde y, p, q) = \Omega (x + \tilde y, \tilde x - 
y, p, q)
\end{equation}
and Eq. (16) is brought to the first heavenly equation [27]

\begin{equation}
\Omega_{\tilde x q} \Omega_{\tilde y p} - \Omega_{\tilde x p} \Omega_
{\tilde y q} = 1
\end{equation}
 
(It is evident that the first heavenly equation can be also obtained when 
other symmetries are assumed, for example

\begin{equation}
\theta '_{\tilde x} = 0 \ \ \ \mbox{and} \ \ \ \theta '_{\tilde y} = 0).
\end{equation}

(b) {\it The second heavenly equation}
 
Here we assume the following symmetry

\begin{equation}
\theta_x = \theta_q \ \ \ \mbox{and} \ \ \ \theta_y = \theta_p
\end{equation}

Consequently, $\theta$ takes the form

\begin{equation}
\theta (x, y, \tilde x, \tilde y, p, q) = \ \Theta (x + q, y + p, \tilde x, 
\tilde y)
\end{equation}
and now Eq. (16) is brought to the second heavenly equation [27]

\begin{equation}
\Theta_{x \tilde x} + \Theta_{y \tilde y} + \Theta_{xx} \Theta_{yy} - 
\Theta^2_{xy} = 0.
\end{equation}

(c) {\it Grant's equation}.

\noindent
Let now

\begin{equation}
\theta_x = \theta_{\tilde x} \ \ \mbox{and} \ \ \theta_{\tilde y} = 0.
\end{equation}

Hence $\theta$ has the form

\begin{equation}
\theta (x, y, \tilde x, \tilde y, p, q) = h (x + \tilde x, y, p, q)
\end{equation}
and Eq. (16) reads

\begin{equation}
h_{xx} + h_{xq} h_{yp} - h_{xp} h_{yq} = 0
\end{equation}

This is Grant's equation [28]

(d) {\it The evolution form of the second heavenly equation}.

Here we assume the symmetry

\begin{equation}
\theta_x = - \theta_{\tilde x} \ \ \mbox{and} \ \ \theta_y = - \theta_q
\end{equation}

Thus

\begin{equation}
\theta (x, y, \tilde x, \tilde y, p, q) = H (x - \tilde x, y-q, \tilde y, p)
\end{equation}
and, consequently, Eq. (16) is reduced to the evolution form of the second 
heavenly equation [29, 30]

\begin{equation}
H_{xx} - H_{y \tilde y} + H_{xy} H_{yp} - H_{xp} H_{yy} = 0
\end{equation}

(e) {\it Husain's equation}

This equation has been found by the reduction of the Ashtekar-Jacobson-
Smolin equations to the $sdiff (\Sigma^2)$ chiral field equations in two 
dimensions [13].
 
In our approach we assume the following symmetry

\begin{equation}
\theta_x = \theta_{\tilde x} \ \ \ \mbox{and} \ \ \ \theta_y = 
\theta_{\tilde y}.
\end{equation}

Therefore
\begin{equation}
\theta (x, y, \tilde x, \tilde y, p, q) = \Lambda (x + \tilde x, y + \tilde y, 
p, q)
\end{equation}
and Eq. (16) takes the form of Husain's equation

\begin{equation}
\Lambda_{xx} + \Lambda_{yy} + \Lambda_{xq} \Lambda_{yp} - \Lambda_{xp} 
\Lambda_{yq} = 0
\end{equation}

To have a contact with some previous works [29, 31, 32] we rewrite Eq. (16) in 
terms of 2-spinors and then, in the differential forms langauge. 
 
To this end put 

\begin{equation}
(x, y, \tilde x, \tilde y, p, q) \equiv (-p^1, -p^2, q_1, q_2, \tilde q_2,
\tilde q_1).
\end{equation}

Hence, Eq. (16) reads now

\begin{equation}
\frac{1}{2} \ \frac{\partial^2 \theta}{\partial p^A \partial\tilde q_B} \ 
\frac{\partial^2 \theta}{\partial p_A \partial\tilde q^B} \ + \ 
\frac{\partial^2 \theta}{\partial p^A \partial q_A} = 0, \ A, B=1, 2, 
\end{equation}
where the spinorial indices are to be manipulated according to the rule

\[
\phi_A = \epsilon_{AB} \phi^B, \ \ \ \phi^A = \epsilon^{BA} \phi_B
\]

\begin{equation}
(\epsilon_{AB}):= \pmatrix{0 & 1\cr -1 & 0\cr} =: (\epsilon^{AB}),  \ \ 
A, B, = 1, 2.
\end{equation}
 
Then the Lagrangian ${\cal L}_\infty$ defined by (15) takes the form

\begin{equation}
{\cal L}_\infty = - \frac{1}{6} \theta \frac{\partial^2 \theta}{\partial p^A 
\partial \tilde q^B} \frac{\partial^2 \theta}{\partial p_A \partial \tilde q_B} 
- \frac{1}{2} \frac{\partial \theta}{\partial p^A} \frac{\partial \theta}
{\partial \tilde q_A}.
\end{equation}

We now denote

\begin{equation}
(\theta_x, \theta_y, \theta_{\tilde x}, \theta_{\tilde y}, \theta_p, \theta_q) 
\ \equiv \ (s_1, s_2, r^1, r^2, \tilde r^2, \tilde r^1).
\end{equation}

Consequently, by straightforward calculations one can show that in terms 
of differential forms the sixdimensional version of the heavenly 
equation (34) can be written as follows

\begin{eqnarray}      
ds^A \wedge d p_A +  dr^A \wedge dq_A + d \tilde r^A \wedge d \tilde q_A 
& = & 0\\ 
(d \tilde q_A \wedge d \tilde q^A + ds^A \wedge d q_A) \wedge d s^B \wedge 
dq_B \wedge dp^C \wedge dp_C & = & 0. \\ 
d p^A \wedge d p_A \wedge d q^B \wedge d q_B \wedge d \tilde q^C \wedge d 
\tilde q_C & \neq & 0
\end{eqnarray}

Thus, to obtain from (38.a, b c) the second heavenly equation one assumes 
(see (21))

\begin{equation}
s_A = \tilde r^A 
\end{equation}

This leads to the equation

\begin{equation}
\frac{1}{2} \ \ \frac{\partial^2 \Theta}{\partial p^A \partial p_B}  \ \ 
\frac{\partial^2 \Theta}{\partial p_A \partial p^B} \ + \ \frac{\partial^2 
\Theta}{\partial p^A \partial q_A} = 0
\end{equation}
\[ \Theta = \Theta (- p^A + \tilde q_A, q_B), \]
which is exactly the second heavenly equation as written in terms of 
2-spinors [29, 31, 32]

To get the first heavenly equation we put

\begin{equation}
s^A + \frac{1}{2} q^A = r^A - \frac{1}{2} p^A.
\end{equation}

Then from (38.a) one infers the existence of a function $\Omega$ such that 

\begin{equation}
\frac{\partial \Omega}{\partial p_A} = \frac{\partial \Omega}{\partial q_A} 
= r^A - \frac{1}{2} p^A = s^A + \frac{1}{2} q^A
\end{equation} 
i.e.,  $\Omega$ is of the form

\begin{equation}
\Omega = \Omega (q_A + p_A, \tilde q_B).
\end{equation}

Consequently, (38.b) and (42) give the first heavenly equation [29, 31, 32]

\begin{equation}
\frac{1}{2} \frac{\partial^2 \Omega}{\partial q^A \partial \tilde q_B} 
\frac{\partial^2 \Omega}{\partial q_A \partial \tilde q^B} + 1 = 0.
\end{equation}
 
Finally, from (41) and (42) we obtain the relation (compare with (17))

\begin{equation}
\theta = \Omega + \frac{1}{2} p^A q_A.
\end{equation}

Analogously one can find other heavenly equations in the spinorial form.

Now it is evident that the sixdimensional version of the second heavenly 
equation (34) implies the following conservation law

\begin{equation}
\frac{\partial}{\partial p_A} \bigg( \frac{1}{2} \frac{\partial^2\theta}{
\partial p^A \partial \tilde q_B} \frac{\partial \theta}{\partial\tilde q^B} 
- \frac{\partial \theta}{\partial q^A} \bigg) = 0 
\end{equation}

According to our philosophy this conservation law overlaps the hierarchy of 
conservation laws for the heavenly equations.

For example, in the case of the second heavenly equation, by (39), we have 
$- \frac{\partial \theta}{\partial p^A} = \frac{\partial \theta}{\partial 
\tilde q_A} $. \ Therefore, in this case (46) gives 

\begin{equation}
\frac{\partial}{\partial p_A} \bigg( \frac{1}{2} \frac{\partial^2 \Theta}
{\partial p^A \partial p_B} \ \frac{\partial \Theta}{\partial p^B} - 
\frac{\partial \Theta}{\partial q^A} \bigg) = 0, 
\end{equation}
where we substituted $\Theta$ in place of $\theta$ (see 40)).

On the other side in the case of the first heavenly equation, employing 
(42) and (45), one quickly finds that the conservation law (46) reads 

\begin{equation}
\frac{\partial}{\partial q_A} \bigg( \frac{\partial^2 \Omega}{\partial q^A 
\partial \tilde q_B} \frac{\partial \Omega}{\partial \tilde q^B} + q_A 
\bigg) = 0.
\end{equation}

It is quite natural to expect that  the conservation law (46) in six 
dimensions generates an infinite hierarchy of conservation laws in four 
dimensions when the heavenly equaitons are assumed to hold. In order to 
prove this statement and also to find the relation of our approach with 
the previous 
works by Boyer and one of us (J.F.P) [29, 32] and by Strachan [33] we should 
first find the general theory of symmmetry reduction of the sixdimensional 
version of the second heavenly equation to the heavenly equations. The work on 
this theory is in progress.

Finally, we are going to consider the Moyal deformation of Eq. (16). To this 
end we consier the SDYM equations (7. a, b) assuming that the potentials 
are now the self-adjoint operator-valued functions on $V\subset R^4$ acting 
in a Hilbert space ${\cal H} = L^2 (R^1)$. Thus we now deal with the equations

\begin{eqnarray}
\partial_{\tilde x} \hat A_{\tilde y} - \partial_{\tilde y} \hat A_{\tilde x} 
+ \frac{1}{i\hbar} [\hat A_{\tilde x}, \hat A_{\tilde y} ] & = & 0\\ 
\partial_x \hat A_{\tilde x} + \partial_y \hat A_{\tilde y} &= &0\\ 
\hat A^+_{\tilde x} = \hat A_{\tilde x}, \hat A^+_{\tilde y} &=&\hat   
A_{\tilde y}. 
\end{eqnarray}

Then from (49.b) we get

\[ \hat A_{\tilde x} = - \partial_y \hat\theta \ \ \mbox{and} \ \ 
\hat A_{\tilde y} = \partial_x \hat \theta, \]

\begin{equation}
\hat \theta = \hat \theta (x, y, \tilde x, \tilde y) = \hat \theta^+.
\end{equation}

Inserting (50) into (49.a) one obtains

\begin{equation}
\partial_x \partial_{\tilde x} \hat \theta + \partial_y \partial_{\tilde y} 
 \hat\theta + \frac{1}{i\hbar} [\partial_x \hat\theta , \partial_y \hat\theta 
 ] = 0.
\end{equation}

Straightforward calculations show that Eq. (51) can be derived from the 
following variational principle 

\[ \delta S^{(q)} = 0, \ \ \ S^{(q)} = \int_V {\cal L}^{(q)} d v \]

\[ {\cal L}^{(q)} := Tr \ \bigg\{ 2\pi \hbar \big[ \frac{-1}{3i\hbar} 
\hat \theta [ \partial_x \hat \theta , \partial_y \hat \theta] + 
\frac{1}{2} \big( (\partial_x \hat \theta)(\partial_{\tilde x} \hat \theta)
+ (\partial_y \hat \theta)(\partial_{\tilde y} \hat \theta ) ] \bigg\} \]

\begin{equation}
= 2 \pi \hbar \sum_j < \psi_j | \bigg\{ \frac{-1}{3 i \hbar} \hat \theta 
[\partial_x \hat\theta , \partial_y \hat \theta ] + \frac{1}{2} \big( 
(\partial_x \hat \theta)(\partial_{\tilde x} \hat\theta) + (\partial_y 
\hat\theta)(\partial_{\tilde y} \hat\theta) \big) \bigg\} | \psi_j >, 
\end{equation}
where $\{ |\psi_j > \}_{j\epsilon {\cal N}}$ constitutes an orthonormal
basis in ${\cal H}$

\begin{equation}
< \psi_j | \psi_k > = \delta_{jk}, \ \ \sum_j | \psi_j ><\psi_j | = \hat 1.
\end{equation}

Employing  the Weyl-Wigner-Moyal formalism [34-40] one can bring ${\cal L}^{(q)}$ 
to the following form

\[ {\cal L}^{(q)} = 2 \pi \hbar \sum_j \int_{R^2} \ \bigg\{ \frac{-1}{3} \  
\theta * \{ \partial_x \theta , \partial_y \theta \}_M + \frac{1}{2} 
\big( (\partial_x \theta) * (\partial_{\tilde x} \theta) \]
\begin{equation}
+ (\partial_y \theta) * (\partial_{\tilde y} \theta) \big) \bigg\} \rho_j 
(p, q) dpdq, 
\end{equation}
where

\begin{equation}
\theta = \theta (x, y, \tilde x, \tilde y, p, q):= \ \int^{+\infty}_{-\infty}
\ < q - \frac{\xi}{2} | \hat\theta | q + \frac{\xi}{2} > \ \ \mbox{exp} \ \ 
\bigg(\frac{ip\xi}{\hbar} \bigg) d \xi 
\end{equation}
(the {\it Weyl correspondence}), and $\rho_j = \rho_j (p, q)$ denotes the 
{\it Wigner function} for $| \psi_j>,$ i.e.,  

\begin{equation}
\rho_j = \rho_j (p, q):= \frac{1}{2\pi\hbar} \ \int^{+\infty}_{-\infty} \ <
q - \frac{\xi}{2} | \psi_j >< \psi_j | q + \frac{\xi}{2} > \ \ 
\mbox{exp} \ \ \bigg( \frac{ip\xi}{\hbar}\bigg) d \xi 
\end{equation}

Moreover, the {\it Moyal $*$-product} is defined by 

\[ * := \ \mbox{exp} \ \bigg( \frac{i\hbar}{2} \stackrel{\leftrightarrow}
{\cal P} \bigg), \]

\begin{equation}
\stackrel{\leftrightarrow}{\cal P} := \ 
\frac{ \stackrel{\leftarrow}{\partial}}{\partial q} 
\frac{ \vec \partial}{\partial p} - 
\frac{ \stackrel{\leftarrow}{\partial}}{\partial p}
\frac{ \vec \partial}{\partial q}, 
\end{equation}
and $\{ . , . \}_M$ stands for the {\it Moyal bracket}

\begin{equation}
\{ f_1, f_2\}_M:= \frac{1}{i\hbar} \ (f_1 * f_2 - f_2 * f_1) = \frac{2}{\hbar} 
f_1 \ \mbox{sin} \ \big(\frac{\hbar}{2} \stackrel{\leftrightarrow}{\cal P} 
\big) f_2, 
\end{equation}
\[ f_1 = f_1 (x, y, \tilde x, \tilde y, p, q), f_2 = f_2 (x, y, \tilde x, 
\tilde y, p, q). \]

From (53) one quickly finds that

\begin{equation}
\sum_j \rho_j = \frac{1}{2\pi\hbar}.
\end{equation}

Inserting (59) into (54) we get the action ${\cal S}^{(q)}$ to be of the form

\[  {\cal S}^{(q)} = \int_{V \times R^2} \ \{\frac{-1}{3} \theta * \{ 
\partial_x \theta , \partial_y \theta \}_M + \frac{1}{2} ( (\partial_x 
\theta) * ( \partial_{\tilde x} \theta)  \]

\begin{equation}
+ (\partial_y \theta )*( \partial_{\tilde y} \theta )) \} d v d p d q.
\end{equation}

This action leads to the Moyal deformation of the sixdimensional version of 
the second heavenly equation

\begin{equation}
\partial_x \partial_{\tilde x} \theta + \partial_y \partial_{\tilde y} 
\theta + \{ \partial_x \theta , \partial_y \theta \}_M = 0.
\end{equation}

As

\begin{equation}
\lim_{\hbar \to 0} f_1 * f_2 = f_1 f_2 \ \mbox{and} \ \ \lim_{\hbar \to 0} 
\{ f_1, f_2\}_M = \{f_1, f_2\}_P
\end{equation}
one quickly finds that the sixdimensional version of the second heavenly 
equation (16) is the $\hbar \to 0$ limit of Eq. (61) and

\begin{equation}
\lim_{\hbar \to 0} {\cal S}^{(q)} = {\cal S}_\infty .
\end{equation}

Gathering all that one concludes tha the following relations hold

${\cal O}$ SDYM EQUATIONS $\stackrel{\hbar \to 0} {\longrightarrow}$ 
HEAVENLY EQUATIONS $\stackrel{N \to \infty}{\longleftarrow} $ su (N) SDYM 
EQUATIONS(where ${\cal O}$ denotes the set of the self-adjoint operators in 
${\cal H}$).

{\bf \it Remark}:

The Lagrangian (10) has been considered by Leznov [41] and then by Parkes 
[42]. Another SDYM Lagrangian has been proposed in [43, 44]. It has been 
written within the J formalism and we have not been able to find the analogy 
of the matrix J in self dual gravity.  We suppose that this analogy can be 
found  when 
the Weyl-Wigner-Moyal formalism is used and then when the $\hbar \to 
0$ limit is considered.   The work on this problem is in progress.  We are 
grateful to the referee for pointing out the papers [41-44]  and for stating 
the question of uniqueress of the $N \to \infty$ limit for the SDYM 
Lagrangians. 

We are indebted to H. Garc\'\i a-Compe\'an for useful discussions.  \ \ 
One of us (M.P.) is grateful to the staff of the Department of Physics at 
CINVESTAV for the warm hospitality.

This work is supported by CONACyT and CINVESTAV, M\'exico, D.F., M\'exico.

\end{document}